\documentclass[aps]{revtex4}
\usepackage{graphicx}
\usepackage{amsmath}
\usepackage{amssymb}

\begin{document}

\title[MOG BH lensing observables in weak and strong field of gravity]{Modified gravity black hole lensing observables in weak and strong field of gravity}

\author{R.N. Izmailov}
\email{izmailov.ramil@gmail.com}
\affiliation{Zel'dovich International Center for Astrophysics, Bashkir State Pedagogical University, 3A, October Revolution Street, Ufa 450008, RB, Russia}
\author{R.Kh. Karimov}
\email{karimov\_ramis\_92@mail.ru}
\affiliation{Zel'dovich International Center for Astrophysics, Bashkir State Pedagogical University, 3A, October Revolution Street, Ufa 450008, RB, Russia}
\author{E.R. Zhdanov}
\email{zhdanov@ufanet.ru}
\affiliation{Zel'dovich International Center for Astrophysics, Bashkir State Pedagogical University, 3A, October Revolution Street, Ufa 450008, RB, Russia}
\author{K.K. Nandi}
\email{kamalnandi1952@yahoo.co.in}
\affiliation{Zel'dovich International Center for Astrophysics, Bashkir State Pedagogical University, 3A, October Revolution Street, Ufa 450008, RB, Russia}
\affiliation{High Energy and Cosmic Ray Research Center, University of North Bengal, Siliguri 734 013, WB, India}
%\date{\today}

\begin{abstract}

We extend a recent work on weak field first order light deflection in the MOdified Gravity (MOG) by comprehensively analyzing the actual observables in gravitational lensing both in the weak and strong field regime. The static spherically symmetric black hole (BH) obtained by Moffat is what we call here the Schwarzschild-MOG (abbreviated as SMOG) containing repulsive Yukawa-like force characterized by the MOG parameter $\alpha>0$ diminishing gravitational attraction. We point out a remarkable feature of SMOG, viz., it resembles a regular \textit{brane-world} BH in the range $-1<\alpha <0$ giving rise to a negative "tidal charge" $Q$ ($=\frac{1}{4}\frac{\alpha }{1+\alpha}$) interpreted as an imprint from the $5D$ bulk with an imaginary source charge $q$ in the brane. The Yukawa-like force of MOG is attractive in the brane-world range enhancing gravitational attraction. For $-\infty <\alpha <-1$, the SMOG represents a naked singularity. Specifically, we shall investigate the effect of $\alpha $ or Yukawa-type forces on the weak (up to third PPN order) and strong field lensing observables. For illustration, we consider the supermassive BH SgrA* with $\alpha =0.055$ for the weak field to quantify the deviation of observables from GR but in general we leave $\alpha$ unrestricted both in sign and magnitude so that future accurate lensing measurements, which are quite challenging, may constrain $\alpha$.
\end{abstract}

\maketitle

%%%%%%%%%%%%%%%%%%%%%%%%%%%%%%%%%%%%  DATE  %%%%%%%%%%%%%%%%%%%%%%%%%%%%%%%%%%%%

\section{Introduction}

Recently, by integrating the Fermat potential in MOdified Gravity (MOG), the
weak field first order light deflection angles caused by a point-like star
and a large compact object have been found and their deviations from those
of general relativity (GR) have been discussed \cite{MR:2019}.
Although light deflection by the gravitational field of an object (lens) is
the core physics behind lensing, it is not directly measured. What one
measures instead is the image position, flux etc. called lensing \textit{%
observables} and their analyses is always a step ahead in understanding
the nature of the lens. While the traditional\ weak field lensing is
adequately described by the Euclidean lens geometry, it does not work in the
strong field regime due to the existence of a photon sphere of the
deflecting lens demarcating the strong field limit. Light rays from the
source get captured on the sphere so that they cannot form images visible to
an external observer. Light rays passing very near to the sphere (at some
minimum impact parameter) make several loops around the sphere before
emerging in another direction forming observable images. Therefore, lensing
behavior could be a useful diagnostic to differentiate between Schwarzschild
black hole of GR (SGR) and of MOG.

The \textit{raison d'\^{e}tre} for MOG is provided by the fact that the GR
postulate of attractive dark matter and repulsive dark energy, invoked to
explain respectively the galactic flat rotation curves and cosmological
acceleration, continues to remain enigmatic. Despite several speculations
(WIMPs, scalar fields, etc), all experiments to date have failed to directly
detect the elusive dark matter \cite{Aprileetal:2012,Akeribetal:2014,Agneseetal:2014}.
This situation has led physicists to devise alternative
theories, classified under the name modified gravity (MOG), that seek to
preserve the successes of GR but do not require the postulate of
undetectable matter. Some notable alternative theories are, but not limited
to, Weyl conformal gravity \cite{MO:2011}, see also \cite{NB:2012},
MOdified Newtonian Dynamics (MOND) \cite{Milgrom:1983} and $f(R)-$%
gravity \cite{NO:2006}. The static spherically symmetric BH
solution obtained by Moffat \cite{Moffat:2006} is what we call here the
Schwarzschild-MOG (abbreviated as SMOG). The SMOG is characterized by a
massive vector field with an enhanced Newtonian acceleration defined by a
gravitational constant $G=G_{N}(1+\alpha )$, where $G_{N}$ is the Newtonian
gravitational constant, $\alpha $ is the MOG parameter representing a
repulsive Yukawa-like force ($\alpha >0$). SMOG leads to SGR at $\alpha =0$.

MOG has already been receiving attention among the astrophysics community\
since it successfully explains not only the local weak field tests, but also
the rotation curves of nearby galaxies \cite{BM:2006a,BM:2006b,MR:2013,PAR:2017a,PAR:2017b}, the dynamics of globular clusters
in the galactic halo \cite{BM:2006a,BM:2006b,BM:2007,MT:2008}, and cosmological observations \cite{MT:2013}
without requiring the dark matter paradigm of GR \cite{PAR:2017a,PAR:2017b,MR:2013,MT:2015,MR:2014,Moffat:2015a,Moffat:2015b,HJ:2015}. Properties of
accretion disks around supermassive BHs in MOG and a detectable influence of
the parameter $\alpha $ on the accretion charateristics of supermassive BH
(SMBH) has been noted \cite{PAR:2017a}. Stability of circular orbits
around a SMOG in the ambience of a weak magnetic field has been studied
\cite{HJ:2015}. The latest success of MOG \citep{MT:2019} is
that it can reproduce with reasonable accuracy the observed velocity
dispersion in the ultra-diffuse galaxy $NGC1052-DF2$ that shows it to be
devoid of dark matter \cite{vanDokkumetal:2018}. Gravitational lensing is
another useful diagnostic \cite{LS:2013,Moffat:2015a,MRT:2018} that is able to distinguish the predictions from competing theories.
Our aim is to theoretically distinguish the lensing manifestations of the
Yukawa-like forces characterized by the MOG parameter $\alpha $.

In this paper, it will be first argued that SMOG resembles a regular
brane-world BH in the interval $-1<\alpha <0$, which joins the SMBH interval
$0<\alpha <2.47$ \cite{PAR:2017a} across the SGR divide $\alpha =0$.
We shall then proceed beyond the first order light deflection \cite{MR:2019} to analyse the actual lensing observables using completely
different methods. For the weak field regime, we shall adopt Keeton-Petters
(KP) method \cite{KP:2006} to see the influence of $\alpha $ on the image positions,
magnifications, centroid and differential time delay between images. For the
strong field regime, we shall adopt Bozza's method \cite{Bozza:2002} to analyse the
angular radius of the BH \textit{shadow} $\theta _{\infty }$, image
separation $s$ and flux ratio $r$, all as functions of $\alpha $. Both the
methods work for a larger interval $-1<\alpha <\infty $ without any
pathology so that the value of $\alpha $ may be left essentially open to be
constrained by future lensing observations. However, for numerical
illustration, we shall assume the SMBH SgrA* as a toy model of SMOG to
numerically estimate the degree of deviation from SGR values within a finite
interval, $-1<\alpha <3$, slightly larger than the one in \cite{PAR:2017a,PAR:2017b}. The obtained numerical values are more indicative than exact since
the mass and distance to the lens are not accurately known,

The paper is organized as follows: We outline in Sec.2 the nature of SMOG
for different $\alpha $. In Sec.3, we work out weak field light deflection up to third order and analyze the weak field
lensing observables using KP method and in Sec.4, the strong field light
deflection and lensing observables using Bozza's method.\textbf{\ }We shall
summarize the obtained results in Sec.5. We take $G_{N}=1$, $c=1$ unless
specifically restored.

\section{The nature of SMOG for different $\alpha$}

The departure of MOG from GR is characterized by Moffat's postulate \cite{Moffat:2015a} that
the gravitational "source charge" $q$ of the vector field $\phi _{\mu }$ is
proportional to the mass $M$ of the gravitating source (since compact
objects including black holes are\ generally electrically neutral) so that
\begin{equation}
q=\pm \sqrt{\alpha G_{N}}M,
\end{equation}%
where $\alpha(=\frac{G-G_{N}}{G_{N}}$) $>0$ determines the strength of gravitational vector forces \cite{MR:2013}. The positive value ($\alpha >0$) produces a repulsive gravitational Yukawa-like force. The simplest case of spherical
symmetry yields a solution%
\begin{equation}
\phi _{0}=-\alpha G_{N}M\left( \frac{\exp (-\widetilde{\mu }r)}{\widetilde{\mu }r}\right),
\end{equation}%
where $\widetilde{\mu }$ is an independent mass parameter. Newtonian gravity
is recovered in the weak field limit $\widetilde{\mu }r<<1$. The SMOG metric
is given by \cite{Moffat:2006}
\begin{eqnarray}
ds^{2} &=&-A(r)c^{2}dt^{2}+B(r)dr^{2}+C(r)\left( d\theta ^{2}+\sin^{2}\theta \phi ^{2}\right), \\
A(r) &=&\frac{1}{B(r)}=-\frac{2G_{N}(1+\alpha )M}{c^{2}r}+\frac{%
G_{N}^{2}M^{2}(1+\alpha )\alpha }{c^{4}r^{2}}, \\
C(r) &=&r^{2}.
\end{eqnarray}%
Redefining the mass $M$ in relativistic units and a dimensionless charge $Q_{0}$ as
\begin{eqnarray}
m_{\bullet } &=&\frac{GM}{c^{2}}=\frac{G_{N}(1+\alpha )M}{c^{2}}, \\
Q_{0} &=&\frac{\alpha }{1+\alpha },
\end{eqnarray}%
we rewrite the SMOG metric (3-5) in the form

\begin{eqnarray}
A(r) &=&1-\frac{2m_{\bullet }}{r}+\frac{Q_{0}m_{\bullet }^{2}}{r^{2}}, \\
C(r) &=&r^{2}.
\end{eqnarray}%
This is the form suitable for using the KP method \cite{KP:2006}. For using Bozza's
method \cite{Bozza:2002}, the scaled metric (8,9) with a radius unit $r_{s}=2m_{\bullet}=%
\frac{2G_{N}(1+\alpha )M}{c^{2}}=1$ is more convenient and is given by

\begin{eqnarray}
A(r) &=&\frac{1}{B(r)}=1-\frac{1}{r}+\frac{Q}{r^{2}}, \\
C(r) &=&r^{2}\text{, }Q=\frac{1}{4}\left( \frac{\alpha }{1+\alpha }\right) .
\end{eqnarray}%
For both the metric forms, we choose $-1<\alpha <3$. The spacetime metric
(8,9) has no central singularity if $\alpha >0$ or $Q_{0}<1$, because of
Yukawa-type repulsion on test particles near $r=0$, while it has a \textit{%
naked singularity} if $Q_{0}>1$, which occurs when $\alpha <-1$. The BH or
naked singularity nature of SMOG for different values of $\alpha $ are shown
in Fig.1.

\begin{figure}
  \centerline{\includegraphics[scale=0.5]{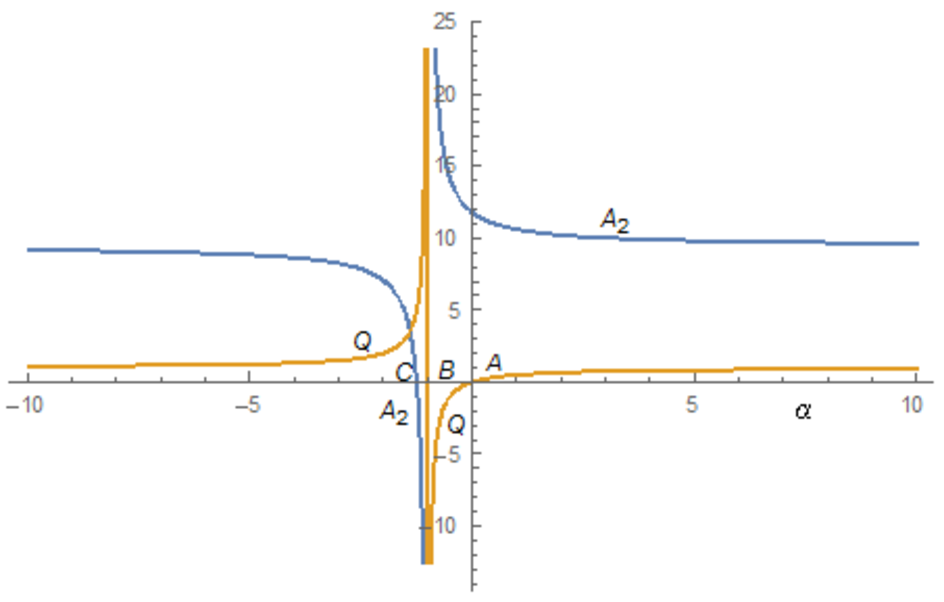}}
  \caption{The nature of solutions depending on the range of the MOG parameter $\protect\alpha $.}
\end{figure}

The spacetime metric form (10,11) resembles a \textit{brane-world BH} for $-1<\alpha <0$ or $Q<0$  \cite{Dadhichetal:2000} belonging to the brane-world theory developed by Sasaki et al. \cite{SSM:2000}. The quantity $Q<0$ has been interpreted as negative "tidal charge" arising out of the free gravitational field propagating in the bulk and its effect in the galactic halo observables has been studied in the literature \cite{Nandi:2009}. The tidal charge that strengthens the gravitational field in the brane as opposed to Yukawa-like repulsion of MOG. Therefore, the left of $A$ in Fig.1 represents brane-world BH valid for $-1<\alpha <0$, while the right side of $A$ represents the SMOG originally conceived, i.e., for $\alpha>0$. The SMOG and the brane-world BH parameters are joined at the SGR divide at $\alpha =0$, hence the BHs are in a sense complementary at the solution level though belonging to entirely different parent theories. We find that the $5D$ bulk imprint $Q<0$ onto the brane should be interpreted as an effect of an \textit{imaginary} source MOG charge $q$ living in the brane. This remarkable complementarity between the SMOG and brane-world BH seems not to have been noticed heretofore.

\section{Weak field light deflection and lensing observables}

The KP method to be used in this section is relatively new and is based on
the PPN expansion of the metric in terms of the gravitational potential of
the lens object. For the SMOG metric (3-5), this is defined by
\begin{equation}
\Phi =-\frac{m_{\bullet }}{r}=-\frac{G_{N}(1+\alpha )M}{c^{2}r}.
\end{equation}%
The expansion yields the weak field light deflection angle for the metric form (8,9) up to \textit{third} PPN\ order in $\left( \frac{m_{\bullet}}{b}\right) $ (see for details, \cite{KP:2006})
\begin{equation}
\widehat{\alpha }_{\text{weak}}(b)=A_{1}\left( \frac{m_{\bullet }}{b}\right)
+A_{2}\left( \frac{m_{\bullet }}{b}\right) ^{2}+A_{3}\left( \frac{m_{\bullet
}}{b}\right) ^{3}+...
\end{equation}%
where $b$ is the impact parameter and $A_{1}$, $A_{2}$, $A_{3}$ are the PPN coefficients. For the metric (8,9) they work out to
\begin{equation}
A_{1}=4,A_{2}=(5-Q_{0})\left( \frac{3\pi }{4}\right) ,A_{3}=\frac{128}{3}%
-16Q_{0},Q_{0}=\frac{\alpha }{1+\alpha }.
\end{equation}

The coefficients (14) may be analyzed by referring to Fig.1. In the right
side of $A$, $\alpha >0$, one has $Q_{0}<1$ and $A_{2}>3\pi $ so that the solution (8,9) represents a non-singular SMOG. At $A,$
where $\alpha =0$, one has $A_{2}=\frac{15\pi }{4}$ and $Q_{0}=0$ so that
one recovers the SGR light deflection up to third order. In the region
between $A$ and $B$, $-1<\alpha <0$, one has $-1<Q_{0}<0$ (interpreted as
tidal charge in the brane) but $A_{2}>3\pi $ so that the solution again
represents a non-singular SMOG in the braneworld sector. At $B$, $\alpha =-1$%
, one has $Q_{0}\rightarrow \infty $, $A_{2}\rightarrow \infty $, the
solution is undefined. In the region between $B$ and $C$, $-\frac{5}{4}%
<\alpha <-1$, one has $A_{2}\leq 0$ and $Q_{0}>1$ so that the SMOG
represents a naked singularity. At $C$, one has $\alpha =-\frac{5}{4}$, $%
A_{2}=0$, $Q_{0}=5$, the solution represents a naked singularity. On the
left side of $C$, $-\infty <\alpha <-\frac{5}{4}$, one has $0<A_{2}<3\pi $
and $Q_{0}>1$, so that the solution again represents a naked singularity.
Thus, in all, the left side of $B$ belongs to the region of naked
singularity.

The KP weak field lensing observables are based on the following algorithm:
Use the lens equation from Euclidean lens geometry with $D=d_{LS}/d_{S}$ [35]
(the angles and distances are shown in Fig.2)
\begin{equation}
\tan \mathfrak{B=}\tan \vartheta -D\left[ \tan \vartheta +\tan \left(
\widehat{\alpha }_{\text{weak}}-\vartheta \right) \right] ,
\end{equation}%
and convert to scaled variables by
\begin{equation}
\beta =\frac{\mathcal{B}}{\vartheta _{E}},\theta =\frac{\vartheta }{%
\vartheta _{E}},\widehat{\tau }=\frac{\tau }{\tau _{E}},\tau _{E}=\frac{%
d_{L}d_{S}}{cd_{LS}}\vartheta _{E}^{2},
\end{equation}%
where $\vartheta _{\bullet }$ is the angular radius of the lens, $\vartheta_{E}$ is the Einstein angle. For a given $\widehat{\alpha }_{\text{weak}}$
determined by a given theory such as in Eq.(13), and a for chosen source angle $\beta$, the unknown image angle $\theta $ is PPN expanded as a series%
\begin{equation}
\theta =\theta _{0}+\theta _{1}\varepsilon +\theta _{2}\varepsilon ^{2}+...
\end{equation}%
where $\varepsilon =\frac{\vartheta _{\bullet }}{\vartheta _{E}}$ is a small
expansion parameter. The expansion is then put back in the lens equation and
by setting to zero the coefficients of the powers of $\varepsilon $, we
obtain the expressions for the corrections $\theta _{1}$, $\theta _{2}$ in
terms of the known source position $\beta$. Once $\theta (\beta )$ thus
determined, the same algorithm applies to magnification $\mu $ defined by
(in unscaled form)
\begin{equation}
\mu \left( \vartheta \right) =\left[ \frac{\sin \mathfrak{B(\vartheta )}}{%
\sin \vartheta }\frac{d\mathfrak{B(\vartheta )}}{d\mathfrak{\vartheta }}%
\right] ^{-1}.
\end{equation}%
Similar expansion holds for the time delay $\widehat{\tau }$ from images.

For actual predictions, we shall convert back in the sequel to the actually
measurable unscaled quantities like observed image positions $\vartheta
^{\pm }$, received fluxes $F^{\pm }=\left\vert \mu ^{\pm }\right\vert F_{%
\text{source}}$ from positive and negative parity images and the
observable differential time delay $\Delta \tau $ between them. We take for
the SMBH SgrA* the measured values $M\simeq 4.3\times 10^{6}M_{\odot
},d_{L}\simeq 8.28$ kpc \cite{Gillessenetal:2009}. Assuming that both the source and the observer
reside at the asymptotically flat regime such that $d_{LS}<<d_{S},d_{L}$ (so
$d_{S}\approx d_{L}$) and with typical distances $d_{LS}\sim 1$ to $100$ pc,
we compute (Note: $1$ rad $=206265$ arcsec $=206265$ $\times 10^{6}$ $\mu $%
arcsec):
\begin{eqnarray}
m_{\bullet } &=&(1+\alpha )\times 6.364\times 10^{11}\text{ cm,} \\
\vartheta _{\bullet } &\simeq &\left( m_{\bullet }/d_{L}\right) \text{ rad}%
=(1+\alpha )\times 5.138\text{ }\mu \text{arcsec}, \\
\vartheta _{E} &=&\left( \sqrt{\frac{4m_{\bullet }d_{LS}}{d_{L}d_{S}}}%
\right) \text{ rad}=0.022\sqrt{1+\alpha }\left( \frac{d_{LS}}{1\text{ pc}}%
\right) ^{1/2}\text{ arcsec,} \\
\varepsilon  &=&\frac{\vartheta _{\bullet }}{\vartheta _{E}}=0.00022\sqrt{%
1+\alpha }\left( \frac{d_{LS}}{1\text{ pc}}\right) ^{-1/2}.
\end{eqnarray}

From the Eqs.(19-22), it is seen that the MOG parameter $\alpha $ has
invaded all quantities including $\varepsilon $ and $\vartheta _{E}$ so that
they deviate from the fixed values of SGR ($\alpha =0$), hence the deviation
is already liable to be observed. Note that for our chosen interval, $%
-1<\alpha <3$, $\varepsilon $ continues to remain small justifying the PPN
expansion. However, the expansion parameter $\varepsilon $, the Einstein
angle $\vartheta _{E}$ become imaginary and mass $m_{\bullet }$ become
negative for $\alpha <-1$, that is, the PPN method cannot accommodate the
naked singularity sector in the SMOG.

\begin{figure}
  \centerline{\includegraphics[scale=0.33]{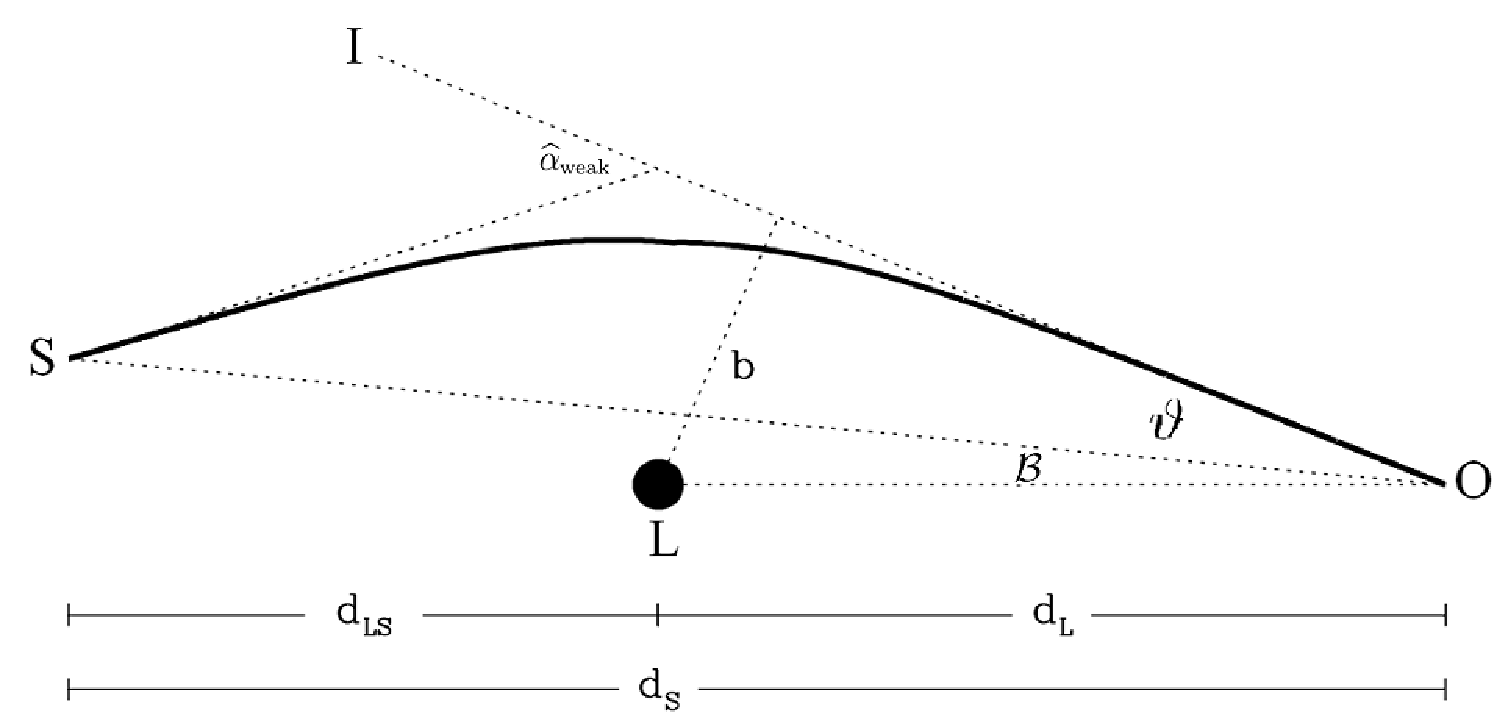}}
  \caption{Weak field lensing geometry with distances as shown. $O$ stands for observer, $L$ stands for lens, $S$ stands for source and $I$ stands for image.}
\end{figure}

Having known the nature of SMOG for different sectors of $\alpha $, we turn
to establish contact with actual observables as a function of the MOG
parameter $\alpha $ for the source angle $\beta \in \lbrack 0,2]$ \ (i.e.,
we are considering sources located only on the positive side of the $OL$
axis for clarity).

\begin{figure}
  \centerline{\includegraphics[scale=0.8]{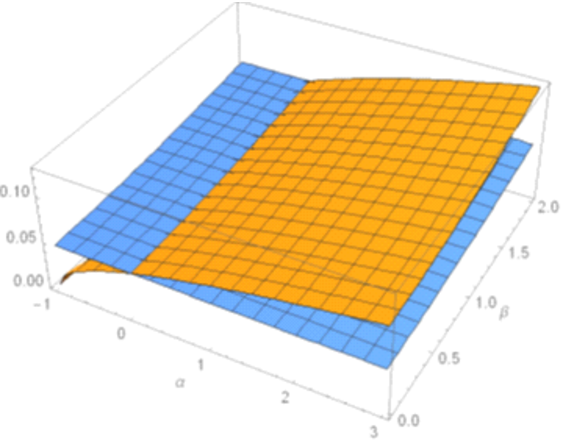}}
  \caption{The comparison of $\protect\vartheta _{\text{tot}}$ in arcsec between the SMOG and SGR as a function of $\protect\alpha $. For easy viewing, the orange colored surface represents SMOG and the blue surface represents SGR.}
\end{figure}

The total angular separation between positive and negative parity images is
defined by $\vartheta _{\text{tot}}\left( \equiv \vartheta ^{+}+\vartheta
^{-}\right)$. Fig.3 illustrates the variation of $\vartheta _{\text{tot}}$
(arcsec) over MOG parameter $\alpha $ and source angle $\beta .$ For SGR ($%
\alpha =0$), it numerically varies between the values $\vartheta _{\text{tot}%
}^{\beta =0}=0.0452$ and $\vartheta _{\text{tot}}^{\beta =2}=0.064$ (blue
surface). In the brane-world sector (orange surface), $\vartheta _{\text{tot}%
}$ continually \textit{decreases} to zero from the SGR values, whereas it
\textit{increases} in the SMOG sector. At $\alpha =0.055$ \cite{MR:2019}, $\vartheta _{\text{tot}}^{\beta =0}=0.0465$ and $\vartheta _{\text{tot%
}}^{\beta =2}=0.0657$ (orange surface), hence the effect of $\alpha$ is to
increase $\vartheta _{\text{tot}}$ by $41.29\%$. The difficulty is that the
absolute magnitudes are too tiny to be measurable. However, if the
source angle $\beta $ is further increased, deviation from SGR becomes quite
prominent.
\begin{figure}
  \centerline{\includegraphics[scale=0.7]{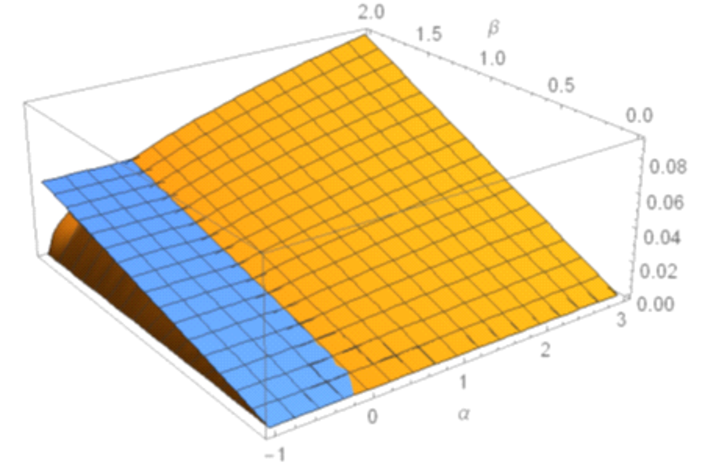}}
  \caption{The comparison of $\Delta \protect\vartheta $ $\left(\equiv \protect\vartheta^{+}-\protect\vartheta^{-}\right) $ in arcsec between the SMOG and SGR as a function of $\protect\alpha$.}
\end{figure}

Fig.4\ illustrates the difference of angles between the positive and
negative parity images, $\Delta \vartheta $ $=\vartheta ^{+}-\vartheta ^{-}$
and the variation of $\Delta \vartheta $ (arcsec) over MOG parameter $\alpha
$ and source angle $\beta .$ For SGR ($\alpha =0$), it varies between the
values $\Delta \vartheta ^{\beta =0}=0.00$ and $\Delta \vartheta ^{\beta
=2}=0.0452$ (blue surface). In the brane-world sector, $\Delta \vartheta $
continually \textit{decreases} to zero from the SGR values, whereas it
\textit{increases} in the SMOG sector. At $\alpha =0.055$ \cite{MR:2019}, $\Delta \vartheta ^{\beta =0}=0.00$ and $\Delta \vartheta ^{\beta
=2}=0.0464$ (orange surface), so the effect of $\alpha $ on the absolute
scale is nearly absent. However, if the source angle $\beta $ is increased
further, deviation from SGR becomes prominent. Incidentally, the plot of the
centroid of the images $\vartheta _{\text{cent}}$ differs very little from
that of $\Delta \vartheta $ above, hence not separately displayed.

Fig.5 illustrates the flux ratio $F_{\text{tot}}/F_{\text{src}}$, where $F_{%
\text{tot}}$($\equiv F^{+}-F^{-}$), is the total flux received at the
observer from the images and $F_{\text{src}}$ is the flux emanated at the
source. Due to light deflection, the two fluxes would be different. The
ratio shows the same surface plot for both SMOG and SGR, meaning that the ratio
is practically \textit{independent} of $\alpha $ for any chosen $\beta \neq 0
$. In other words, SMOG displays the same behavior as that of SGR, once the
source direction $\beta $ is fixed. Hence the flux ratio cannot be used as a
diagnostic for distinguishing between the two BHs.
\begin{figure}
  \centerline{\includegraphics[scale=0.8]{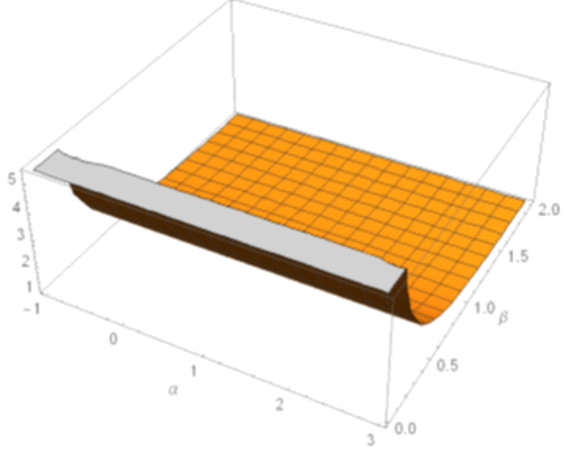}}
  \caption{The comparison of the flux ratio $F_{\text{tot}}/F_{\text{src}}$, where $F_{\text{tot}}$ is the total flux received from the images and $F_{\text{src}}$ is the flux emanated at the source.}
\end{figure}

\begin{figure}
  \centerline{\includegraphics[scale=0.8]{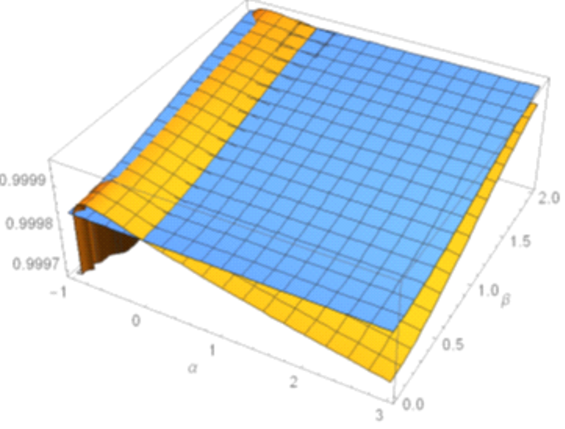}}
  \caption{Comparison of $\Delta F/F_{\text{src}}$ between the SMOG (orange) and SGR (blue).}
\end{figure}

Fig.6 shows comparison of $\Delta F/F_{\text{src}}$ between the SMOG
(orange) and SGR (blue). Practically, the ratio is $\sim 1$ and the effect
of $\alpha $ is not pronounced. Hence the flux ratio is not useful either
for distinguishing the two BHs with current technology.

\begin{figure}
  \centerline{\includegraphics[scale=0.7]{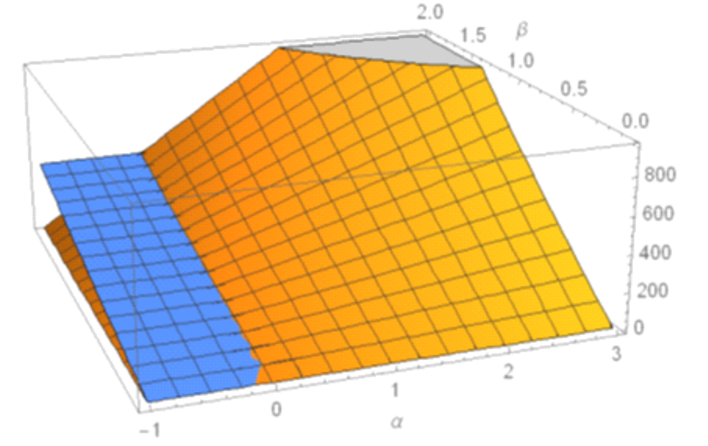}}
  \caption{Differential time delay $\Delta \protect\tau $ from the images.}
\end{figure}

Fig.7 illustrates that the differential time delay $\Delta \tau $ between
two images. It is smaller than that for SGR in the brane-world range ($%
-1<\alpha <0$). The delay rapidly increases from the SGR value (blue
surface) in the SMOG range ($0<\alpha <3$). For instance, it increases from $%
389.69$ seconds ($\alpha =0,\beta =2$) to $411.125$ seconds ($\alpha
=0.055,\beta =2$), and observations should distinguish them.

\section{Strong field light deflection and lensing observables}

The weak field deflection angle $\widehat{\alpha }_{\text{weak}}$ in general
has a major difference with strong field deflection angle $\widehat{\alpha }%
_{\text{strong}}$ since in the latter case the impact parameter $b$ is
\textit{closest} to the photon sphere demarcating the strong field limit,
where the rays suffer a logarithmic divergence or capture so that no
observable images can form. This fact prevents the exact strong deflection
angle to be PPN expanded near photon sphere to yield the same weak field
light deflection for the same $b$. For instance, for the SGR of mass $M$
\citep{IP:2007}%
\begin{eqnarray}
\widehat{\alpha }_{\text{weak}}(b^{\prime }) &=&\frac{4}{3\sqrt{3}}\left(
1-b^{\prime })+O(1-b^{\prime }\right) ^{2}, \\
\widehat{\alpha }_{\text{strong}}(b^{\prime }) &=&-\pi +\log \left[ \frac{%
216(7-4\sqrt{3})}{b^{\prime }}\right] +O(b^{\prime }),
\end{eqnarray}%
where the redefined common impact parameter $b^{\prime }$ is\ $1-b^{\prime }=%
\frac{3\sqrt{3}M}{b}$. At the photon sphere itself, $b=3\sqrt{3}M$ or $%
b^{\prime }=0$, so that one obtains the known result $\widehat{\alpha }_{%
\text{weak}}=\frac{4M}{b}$, whereas $\widehat{\alpha }_{\text{strong}%
}\rightarrow \infty $, as expected. The purpose of this comparison is to
highlight the fundamental differences expected between the set of strong and
weak field lensing observables.

Strong field deflection happens in the vicinity of the photon sphere defined
by the equation [54,55]
\begin{equation}
\frac{C^{\prime }(r)}{C(r)}=\frac{A^{\prime }(r)}{A(r)}
\end{equation}%
which is assumed to admit at least one positive root and the largest root is
called the radius of the photon sphere $r_{m}$. The strong field expansion
will take the photon sphere radius as the starting point, which is required
to exceed the horizon radius of a BH. The minimum impact parameter $u_{m}$
is defined by%
\begin{equation}
u_{m}=\sqrt{\frac{C(r_{m})}{A(r_{m})}}.
\end{equation}

For the calculation of lensing observables, note that the angular separation
of the image from the lens is $\tan \theta =\frac{u}{d_{L}}$, where $d_{L}$
is the distance between the observer and the lens (see Fig.2). Specializing
to the photon sphere $r_{0}=r_{m}$, the \textit{strong field deflection angle%
} proposed by Bozza \citep{Bozza:2002} is
\begin{eqnarray}
\widehat{\alpha }_{\text{strong}}(\theta ) &=&-\overline{a}\log \left( \frac{%
u}{u_{m}}-1\right) +\overline{b}\text{, } \\
u &\simeq &\theta d_{L}\text{ (assuming small }\theta \text{),}
\end{eqnarray}%
where the coefficients $\overline{a}$ and $\overline{b}$ are functions of
the metric coefficients and their derivatives calculated at $r=r_{m}$. The method works so long as $A(r)\neq 1$ \cite{Tsukamoto:2016} which in the case in our paper. The
impact parameter $u_{m}$ is related to the angular separation $\theta $ of
images by the relationship given in Eq.(28). Using this, Bozza proposed
three strong field lensing observables as
\begin{eqnarray}
\theta _{\infty } &=&\frac{u_{m}}{d_{L}}, \\
s &=&\theta _{1}-\theta _{\infty }=\theta _{\infty }\exp \left( \frac{\bar{b}%
}{\bar{a}}-\frac{2\pi }{\bar{a}}\right) , \\[0.5em]
r &=&2.5\log _{10}\left[ \ \exp \left( \frac{2\pi }{\bar{a}}\right) \ \right],
\end{eqnarray}%
where $\theta _{1}$ is the angular position of the outermost image, $\theta
_{\infty }$ ($\mu$arcsec) is the \textit{angular radius of the BH\ shadow} \cite{LS:2013} determined
by the asymptotic position of a set of images in the limit of a large number
of loops the rays make around the photon sphere, $s$ ($\mu$arcsec) is the angular
separation between the outermost image resolved as a single image and the
set of other asymptotic images, all packed together and $r$ (magnitude) is the ratio
between the flux of the first image and the flux coming from all the other
images.

We use the scaled metric form (10,11) together with Eq.(25) which yields the
scaled radius $r_{m}$ of the photon sphere as%
\begin{eqnarray}
r_{m} &=&\left( \frac{3}{4}\right) \left[ 1+\sqrt{1-\frac{32Q}{9}}\right]  \\
&=&\left( \frac{1}{4}\right) \left[ 3+\sqrt{\frac{9+\alpha }{1+\alpha }}%
\right] .
\end{eqnarray}%
The radical sign in (33) implies that the adimensional number $r_{m}$ is
defined for $-\infty <\alpha <-9$ and $-1<\alpha <0$. The first range is
discarded since after restoring the unit, the length $\left[ r_{m}\times
\frac{2G_{N}(1+\alpha )M}{c^{2}}\right]$ cm becomes negative since $\alpha
<-1$. As depicted in Fig.1, this range corresponds to naked singularity and
photon sphere does not exist. The second range corresponds to the
brane-world model with positive $r_{m}$.

The adimensional minimum impact parameter $u_{m}$ is calculated using
Eq.(26) for SgrA*. The outermost image appears at an angle where $\widehat{%
\alpha }_{\text{strong}}(\theta )$ falls below $2\pi $ or when $-\overline{a}%
\log \left( \frac{u}{u_{m}}-1\right) +\overline{b}<2\pi .$ Inserting the
value of $u_{m}$ from Eq.(26), it can be verified that the preceding
inequality is satisfied for $u-u_{m}=0.004$, practically for the whole range
of $\alpha $. We shall use it to plot $\widehat{\alpha }_{\text{strong}%
}(\theta )$ in Fig.8 and the variation of other observables with MOG
parameter $\alpha $ as plotted in Figs.9-12.

\begin{figure}
  \centerline{\includegraphics[scale=0.66]{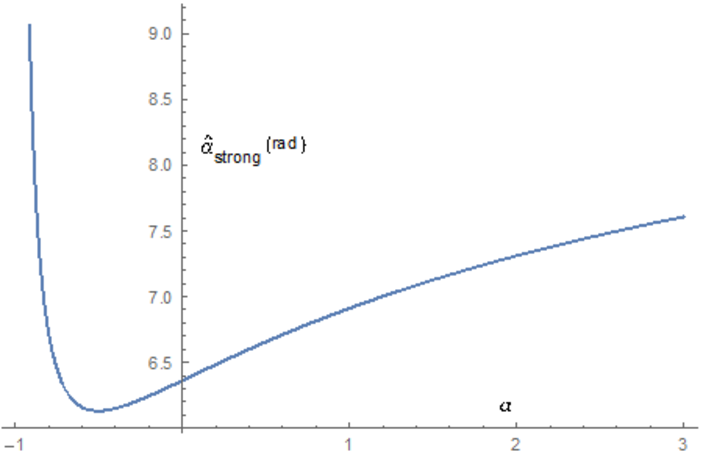}}
  \caption{Light deflection $\widehat{\protect\alpha }_{\text{strong}}(\protect\theta )$ in the strong field limit. It can be seen that the deflection decreases from the SGR ($\protect\alpha =0$) value $6.36$ (rad) and then indefinitely increases as $\protect\alpha \rightarrow -1$ in the brane-world sector\ ($-1<\protect\alpha <0$). On the right of $\protect\alpha =0$, deflection moderately increases to $7.60$ (rad), say at $\protect\alpha =3$, in the SMOG sector.}
\end{figure}

\begin{figure}
  \centerline{\includegraphics[scale=0.66]{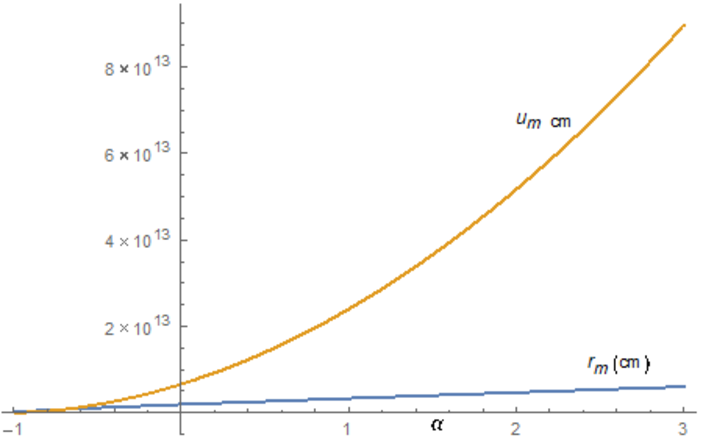}}
  \caption{Radius of the photon sphere $r_{m}$(cm) and the minimum impact parameter $u_{m}$(cm), restoring actual distances in both.}
\end{figure}

\begin{figure}
  \centerline{\includegraphics[scale=0.66]{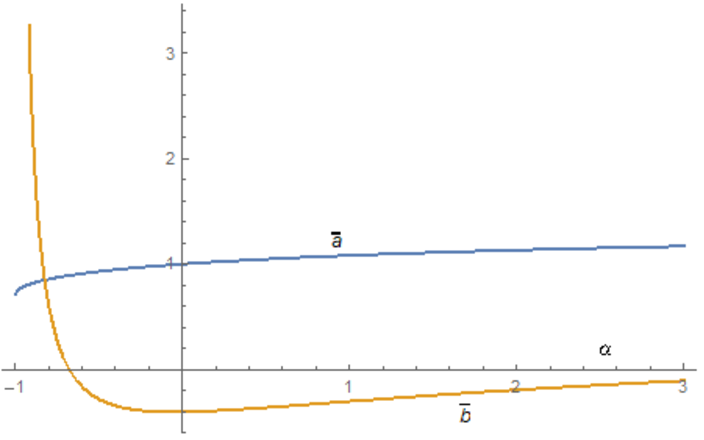}}
  \caption{The strong deflection Bozza coefficients $\overline{a}$, $\overline{b}$ as a function of $\protect\alpha $. The divergence in $\overline{b}$ at the extreme limit $\protect\alpha \rightarrow -1$ causes $\widehat{\protect\alpha }$ $_{\text{strong}}\rightarrow \infty $.}
\end{figure}

\begin{figure}
  \centerline{\includegraphics[scale=0.66]{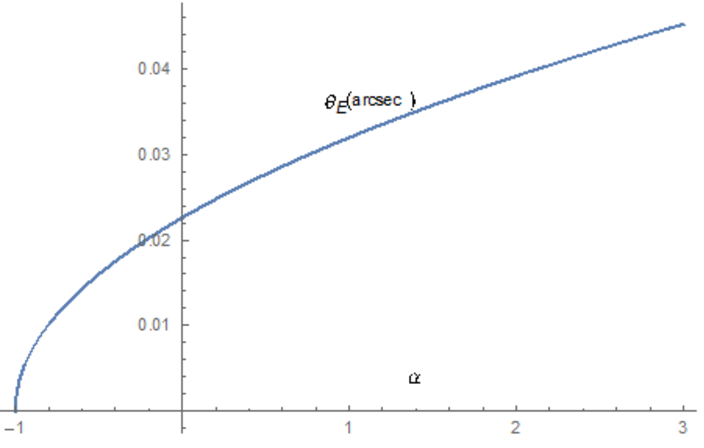}}
  \caption{Angular Einstein radius $\protect\theta _{E}$ (arcsec) as a function of the MOG parameter $\protect\alpha $. The angle decreases to zero in the extreme limit $\protect\alpha\rightarrow -1$.}
\end{figure}

\begin{figure}
  \centerline{\includegraphics[scale=0.66]{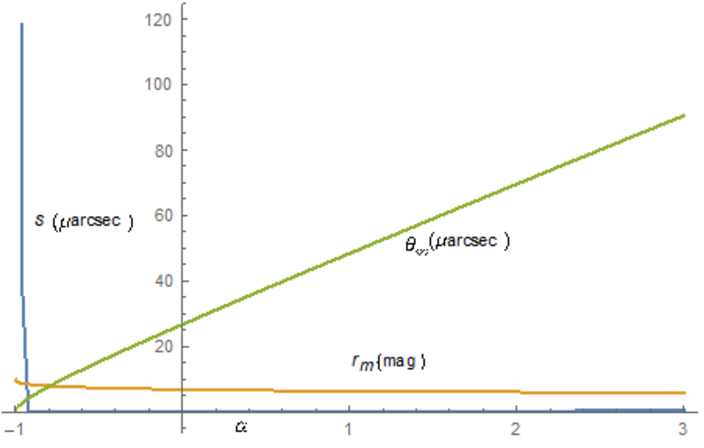}}
  \caption{The strong lensing observables ($s,r,\protect\theta _{\infty }$) all as a function of the MOG parameter $\protect\alpha $. Among the three, the measurement of the angular radius $\protect\theta _{\infty }$ ($\protect\mu $arcsec) of the SgrA* BH shadow is the prime target of the planned EHT. Any observed deviation from the SGR value $26.70$ $\protect\mu $arcsec would put a constraint on $\protect\alpha $.}
\end{figure}

\section{Summary}

Recent works \cite{PAR:2017a,PAR:2017b,MR:2019,MT:2019} have motivated us to comprehensively look into the SMOG from another
angle, viz., the weak and strong field lensing phenomena that are
fundamentally different from each other. The purpose was to explore
observable differences between the lensing properties of SMOG caused by $%
\alpha \neq 0$ and SGR\ ($\alpha =0$). By the same token, the foregoing work
reveals beautiful lensing manifestations of the Yukawa-like forces
determined by the sign of the MOG parameter $\alpha $. In the brane-world
sector ($-1<\alpha <0$), enhancement of the Yukawa-like force is caused by
the bulk effect through a "tidal" charge $Q<0$, while in the SMOG sector $%
\alpha >0$, reduction of the Yukawa-like force is caused by $Q>0$.

We began by pointing out that the SMOG, originally defined for $\alpha >0$,
intriguingly describes a brane-world BH for $-1<\alpha <0$, represented by
the left side of the SGR divide at $\alpha =0$, while naked singularities
appear for $\alpha \leq -1$. These regimes are well demarcated in Fig.1.
Next, we assumed that SgrA* is adequately modeled by the SMOG, both in the
weak and strong field regimes in the range $-1<\alpha <3$ that
combines the sectors of brane-world BH ($-1<\alpha <0$) and SMOG ($0<\alpha <3$), across the divide SGR ($\alpha =0$). We
intentionally chose the combined range to be slightly larger than the
interval suggested by P\'{e}rez et al. \cite{PAR:2017a} for SMBH since we wanted to
accommodate the lensing predictions of all the categories of BHs in a
single framework. However, it turns out that photon sphere does not exist
for the sector $\alpha \leq -1$ (naked singularity). Further, Keeton-Petters
and Bozza observables also throw up divergences for $\alpha \leq -1$, hence
naked singularity cannot be handled by either method although
lensing properties of naked singularity using different approaches have been investigated in the literature
\cite{VE:2000,Virbhadra:2009,ERT:2002,CVE:2001,KS:2003,MM:2005,Whisker:2005,VNC:1998,CH:2018}.
Due to the current inaccuracies in lens mass and distance
measurement, the lensing predictions in this paper should be taken only as
rough estimates since more accurate analyses of the observed data would be
far more complicated.

In the weak field regime, it is evident from Figs.3-6 that the deviations of
observables between SMOG and SGR are too minute to be detectable with
current level of technology within the prescribed interval for MOG parameter
$\alpha $ and source position $\beta $. In the brane-world sector, the
absolute values of all observables decrease with $\alpha $ making any
detection even more difficult. The only possible detectable observable is
the differential time delay $\Delta \tau $, plotted in Fig.7, that is
smaller than that for SGR value in the brane-world range ($-1<\alpha <0$),
but increases in the SMOG sector, say,  from $389.69$ seconds ($\alpha
=0,\beta =2$) to $411.12$ seconds ($\alpha =0.055,\beta =2$).

In the strong field regime, interesting signatures appear. We find that the
deflection $\widehat{\alpha}_{\text{strong}}$ in the brane-world sector\ ($%
-1<\alpha <0$) first decreases from the SGR value $6.36$ rad at $\alpha =0$,
reaches a minimum at $\alpha =-\frac{1}{2}$ and then indefinitely increases
in the extreme limit $\alpha \rightarrow -1$. On the right of $\alpha =0$,
deflection moderately increases to $7.60$ rad at $\alpha =3$ in the SMOG
sector (Fig.8). Fig.9 shows that the separation between the radius of the
photon sphere $r_{m}$ (cm) and the minimum impact parameter $u_{m}$(cm)
tends to be point-like at the limit $\alpha \rightarrow -1$, which means that
the light rays are gradually \textit{forced} into a point-like sink, $\widehat{\alpha }%
_{\text{strong}}\rightarrow \infty $, as reflected also in Fig.8. The root
of this divergence lies in the Bozza coefficient $\overline{b}\rightarrow
\infty $ (Fig.10). There would be no images in this extreme case. (This
behavior is in contrast with the weak field potential $\Phi $ which actually
vanishes) in the limit $\alpha \rightarrow -1$. Also in this limit, $s,\theta _{E}\rightarrow 0$ consistent with point-like sink behavior (Figs.11,12).
The shadow ("silhouette") of SgrA* with an angular radius $\theta _{\infty
}\sim 26.70$ $\mu $arcsec (SGR value, $\alpha =0$) increases almost linearly
with $\alpha >0$ (Fig.12) consistent with the conclusion by Moffat \cite{Moffat:2015a,Moffat:2015b}, but
decreases to\ zero at the extreme limit $\alpha \rightarrow -1$ (Fig.12). If
we adhere to the weak field value $\alpha =0.055$ \cite{MR:2019}, $%
\theta _{\infty }$ increases to $27.92$ $\mu \text{arcsec}$, an increase of
only $4.57\%$ from the SGR value. On the other hand, if we take $\alpha =2.47
$ \cite{PAR:2017a}, then $\theta _{\infty }\sim 79.43$ $\mu $arcsec,
which is an increase of $197.49\%$ from the SGR value. Any measured value of
shadow radius between these limts would in principle constrain the MOG
parameter $\alpha $.

Our broad conclusion is that in the weak field it seems difficult to distinguish between observables in SMOG and SGR in the interval $-1<\alpha <3
$ (applicable for SMBH SgrA*) with the current level of resolution. The strong field observables seem more useful, the most prominent one is the angular radius of the
shadow $\theta _{\infty }$. For this purpose, the Event Horizon Telescope (EHT) based on Very Long Baseline Interferometry (VLBI) is the most
promising project underway. New baselines have been added to achieve an astonishing resolution of $15$ $\mu $arcsec at $345$ GHz \cite{LS:2013}. For recent developments and reviews of experiments, see \cite{BLR:2011,Falcke:2017,Roelofsetal:2017,GP:2018,DeLaurentisetal:2017,Doeleman:2017,Medeirosetal:2018,
Johannsenetal:2016a,Johannsenetal:2016b,Johannsen:2016,LS:2013}.

\section*{Acknowledgements}
% Entry for the table of contents, for this guide only
\addcontentsline{toc}{section}{Acknowledgements}

The reported study was funded by RFBR according to the research Project No. 18-32-00377.

%%%%%%%%%%%%%%%%%%%%%%%%%%%%%%%%%%%%%%%%%%%%%%%%%%

%%%%%%%%%%%%%%%%%%%% REFERENCES %%%%%%%%%%%%%%%%%%

% The best way to enter references is to use BibTeX:

%\bibliographystyle{mnras}
%\bibliography{example} % if your bibtex file is called example.bib

\begin{thebibliography}{99}

\bibitem{MR:2019}
Moffat J.W., Rahvar S., 2019, Mon. Not. R. Astron. Soc., \textbf{482}, 4514

\bibitem{Agneseetal:2014}
Agnese R. et al. (SuperCDMS Collaboration), 2014, Phys. Rev. Lett., \textbf{112}, 241302

\bibitem{Akeribetal:2014}
Akerib D.S. et al. (LUX Collaboration), 2014, Phys. Rev. Lett., \textbf{112}, 091303

\bibitem{Aprileetal:2012}
Aprile E. et al., 2012, Phys. Rev. Lett., \textbf{109}, 181301

\bibitem{MO:2011}
Mannheim P.D.,O'Brien J.G., 2011, Phys. Rev. Lett., \textbf{106}, 121101

\bibitem{NB:2012}
Nandi K.K., Bhadra A., 2012, Phys. Rev. Lett., \textbf{109}, 079001

\bibitem{Milgrom:1983}
Milgrom M., 1983, Astrophys. J., \textbf{270}, 365

\bibitem{NO:2006}
Nojiri S., Odintsov S.D., 2006, Phys. Rev. D, \textbf{74}, 086005

\bibitem{Moffat:2006}
Moffat J.W., 2006, J. Cosmol. Astropart. Phys., 03, 004

\bibitem{BM:2006a}
Brownstein J.R., Moffat J.W., 2006a, Astrophys. J., \textbf{636}, 721

\bibitem{BM:2006b}
Brownstein, J.R., Moffat J.W., 2006b, Mon. Not. R. Astron. Soc., 367, 527

\bibitem{MR:2013}
Moffat J.W.,Rahvar S., 2013, Mon. Not. R. Astron. Soc., \textbf{436}, 1439

\bibitem{PAR:2017a}
P\'{e}rez D., Armengol F.G.L., Romero G.E., 2017a, Phys. Rev. D, \textbf{95}, 104047

\bibitem{PAR:2017b}
Armengol F.G.L., Romero G.E., 2017b, Gen. Relativ. Gravit., \textbf{49}, 27

\bibitem{BM:2007}
Brownstein J.R., Moffat J.W., 2007, Mon. Not. R. Astron. Soc., \textbf{382}, 29

\bibitem{MT:2008}
Moffat, J.W. \& Toth, V.T. 2008, Astrophys. J. \textbf{680}, 1158

\bibitem{MT:2013}
Moffat, J.W. \& Toth, V.T. 2013, Galaxies, \textbf{1}, 65

\bibitem{MT:2015}
Moffat, J.W. \& Toth, V.T. 2015, Phys. Rev. D \textbf{91}, 043004

\bibitem{MR:2014}
Moffat, J.W. \& Rahvar, S. 2014, Mon. Not. R. Astron. Soc. \textbf{441}, 3724

\bibitem{Moffat:2015a}
Moffat, J.W. 2015a, Eur. Phys. J. C \textbf{75}, 175

\bibitem{Moffat:2015b}
Moffat, J.W. 2015b, Eur. Phys. J. C \textbf{75}, 130

\bibitem{HJ:2015}
Hussain, S. \& Jamil, M. 2015, Phys. Rev. D \textbf{92}, 043008

\bibitem{MT:2019}
Moffat J.W., Toth V.T., 2019, Mon. Not. R. Astron. Soc., \textbf{482}, L1

\bibitem{vanDokkumetal:2018}
van Dokkum, P. et al. 2018, Nature (London), \textbf{555}, 629

\bibitem{LS:2013}
Lacroix, T. \& Silk, J. 2013, Astron. Astrophys. \textbf{554}, A36

\bibitem{MRT:2018}
Moffat, J.W., Rahvar, S. \& Toth, V.T. 2018, Galaxies \textbf{6}, 43

\bibitem{KP:2006}
Keeton, C.R. \& Petters, A.O. 2006, Phys. Rev. D \textbf{73}, 044024

\bibitem{Bozza:2002}
Bozza, V. 2002, Phys. Rev. D \textbf{66}, 103001

\bibitem{Dadhichetal:2000}
Dadhich, N. et al. 2000, Phys. Lett. B \textbf{487}, 1

\bibitem{SSM:2000}
Sasaki, M., Shiromizu, T. \& Maeda, K-i. 2000, Phys. Rev. D \textbf{62}, 024008

\bibitem{Nandi:2009}
Nandi, K.K. et al. 2009, Mon. Not. R. Astron. Soc. \textbf{399}, 2079

\bibitem{Gillessenetal:2009}
Gillessen, S. et al. 2009, Astrophys. J. \textbf{707}, L114

\bibitem{IP:2007}
Iyer, S.V. \& Petters, A.O. 2007, Gen. Relativ. Gravit. \textbf{39}, 156

\bibitem{Tsukamoto:2016}
Tsukamoto, N. 2016, Phys. Rev. D \textbf{94}, 124001

\bibitem{VE:2000}
Virbhadra, K.S. \& Ellis, G.F.R. 2000, Phys. Rev. D \textbf{62}, 084003

\bibitem{Virbhadra:2009}
Virbhadra, K.S. 2009, Phys. Rev. D \textbf{79}, 083004

\bibitem{ERT:2002}
Eiroa, E.F., Romero, G. E. \& Torres, D.F. 2002, Phys. Rev. D \textbf{66}, 024010

\bibitem{CVE:2001}
Claudel, C-M., Virbhadra, K.S. \& Ellis, G.F.R. 2001, J. Math. Phys. \textbf{42}, 818

\bibitem{KS:2003}
Kar, S. \& Sinha, M. 2003, Gen. Relativ. Gravit. \textbf{35}, 1775

\bibitem{MM:2005}
Majumdar, A.S. \& Mukherjee, N. 2005, Mod. Phys. Lett. A \textbf{20}, 2487

\bibitem{Whisker:2005}
Whisker, R. 2005, Phys. Rev. D \textbf{71}, 064004

\bibitem{VNC:1998}
Virbhadra, K.S., Narasimha, D. \& Chitre, S.M. 1998, Astron. Astrophys. \textbf{337}, 1

\bibitem{CH:2018}
Cunha, P.V.P. \& Herdeiro, C.A.R. 2018, Gen. Relativ. Gravit. \textbf{50}, 42

\bibitem{BLR:2011}
Broderick, A.E., Loeb, A. \& Reid, M.J. 2011, Astrophys. J. \textbf{735}, 57

\bibitem{Falcke:2017}
Falcke, H. 2017, Journal of Physics: Conf. Ser. \textbf{942}, 012001

\bibitem{Roelofsetal:2017}
Roelofs, F. et al. 2017, Astrophys. J. \textbf{847}, 1

\bibitem{GP:2018}
Giddings, S.B. \& Psaltis, D. 2018, Phys. Rev. D \textbf{97}, 084035

\bibitem{DeLaurentisetal:2017}
De Laurentis, M. et al. 2017, Journal of Physics: Conf. Ser. \textbf{942} 012007

\bibitem{Doeleman:2017}
Doeleman, S.S. 2017, Nature Astronomy (Comment) \textbf{1}, 646

\bibitem{Medeirosetal:2018}
Medeiros, L. et al. 2018, Astrophys. J. \textbf{864}, 7

\bibitem{Johannsen:2016}
Johannsen, T. 2016, Class. Quantum Grav. \textbf{33}, 113001

\bibitem{Johannsenetal:2016a}
Johannsen, T. et al. 2016a, Phys. Rev. Lett. \textbf{117}, 091101

\bibitem{Johannsenetal:2016b}
Johannsen, T. et al. 2016b, Phys. Rev. Lett. \textbf{116}, 031101

\bibitem{Claudel:34}
C.-M. Claudel, K.S. Virbhadra and G.F.R. Ellis, J. Math. Phys. \textbf{42}, 818--838 (2001).

\bibitem{Virbhadra:35}
K.S. Virbhadra and G.F.R. Ellis, Phys. Rev. D \textbf{65}, 103004 (2002).

\end{thebibliography}

% Alternatively you could enter them by hand, like this:

%%%%%%%%%%%%%%%%%%%%%%%%%%%%%%%%%%%%%%%%%%%%%%%%%%

\end{document}